\documentclass[preprint,aps,superscriptaddress,nofootinbib,
tightenlines,floats,floatfix]{revtex4}

\newcommand{\beq}{\begin{equation}}
\newcommand{\eeq}{\end{equation}}
\newcommand{\beqa}{\begin{eqnarray}}
\newcommand{\eeqa}{\end{eqnarray}}

\newcommand{\no}{\nonumber}
\def\lsim{\lesssim}
\def\gsim{\gtrsim}

\begin{document}

\preprint{\vbox{
  \hbox{SCIPP--02--09} \hbox{WIS/24/02--Jun--DPP} }}

\title{\boldmath Product Groups, Discrete Symmetries, and Grand Unification}

\vspace*{1.5cm}

\author{Michael Dine\vspace{10pt}}
\affiliation{Santa Cruz Institute for Particle Physics\\
     Santa Cruz CA 95064, USA\\
dine@scipp.ucsc.edu\\
\vspace{6pt}}

\author{Yosef Nir}
\affiliation{Department of Particle Physics, Weizmann Institute of Science\\
        Rehovot 76100, Israel\\
yosef.nir@weizmann.ac.il\\
$\phantom{}$}

\author{Yael Shadmi\footnote{Incumbent of a Technion Management
Career Development Chair}}
\affiliation{Department of Physics, Technion\\
        Haifa 32000, Israel\\
yshadmi@physics.technion.ac.il\\
$\phantom{}$}

\begin{abstract}
We study grand unified theories based on an $SU(5)\times SU(5)$
gauge group in which the GUT scale, $M_{\rm GUT}$, is the VEV of an exact or
approximate modulus, and in which fast proton decay is avoided
through a combination of a large triplet mass and small
triplet couplings. These features are achieved by discrete
symmetries. In many of our models, $M_{\rm GUT}$ is generated
naturally by the balance of higher dimension terms that lift
the GUT modulus potential, and soft supersymmetry breaking masses.
The theories often lead to interesting patterns of quark and
lepton masses. We also discuss some distinctions between grand unified
theories and string unification.
\end{abstract}

\maketitle

%%%%%%%%%%%%%%%%%%%%%%%%%%%%%%%%%%%%%%%%%%%
%%%%%%%%%%%%%%%%%%%%%%%%%%%%
\section{Introduction}
\label{intro}
%%%%%%%%%%%%%%%%%%%%%%%%%%%%
%%%%%%%%%%%%%%%%%%%%%%%%%%%%%%%%%%%%%%%%%%%
Lacking direct evidence, the unification of couplings
is one of the few experimental hints both of supersymmetry and
of unification of the gauge interactions of the standard model.
The separation of the unification scale from the Planck scale
has long suggested that the methods of effective field theory
might be appropriate in understanding unification.
Probably the most troubling puzzle of such grand unified field
theories (GUTs) is why Higgs doublets should be light,
while their colored partners are massive.

String theory offers an interesting perspective on these issues.
In string theories, one often has unification of couplings, even
though there is typically no scale at which one can speak of a
four dimensional unifying gauge group.  The unification scale,
instead, is to  be identified with the radius of some internal
space, or with the string scale.  In most cases, this scale is the
threshold, not for a finite number of states, but for an infinite
number of states.  One of the appealing features of string models,
stressed even in textbooks \cite{Green:sp}, is that they can
readily produce light doublets without light triplets.

String theory also sheds light on some of the traditional
questions of model building.  In particular, it was long suspected
that one should not expect continuous global symmetries in a
quantum theory of gravity, and indeed there are no such symmetries
in string theory \cite{Banks:1988yz}.  String theory does frequently yield rather
intricate patterns of discrete symmetries.

This discussion suggests that we make a distinction between {\it
Grand Unification}, i.e. theories where, for some range of energy
scales, there are a finite number of fields and the gauge groups
of the standard model are unified in a larger group, and {\it
String Unification}, where the couplings are unified, but there is
no such range of energies.  There are a number of reasons to
pursue Grand Unification, even if one imagines that the underlying
theory is string theory.  Perhaps the most important of these is
coupling unification itself.  Coupling unification is a robust
property of the weakly coupled heterotic string.   It is also true
of theories in which some larger gauge symmetry is broken by
Wilson lines (e.g. strongly coupled heterotic string, $G_2$
compactifications, etc.).  But it does not hold more generally,
and it is thus not clear in what sense it is a {\it prediction} of
string theory.  Certainly one could imagine that the explanation
of coupling unification, even in string theory, is Grand
Unification.  Second, even though the appearance of massless Higgs doublets
is an impressive feature of string theory, the absence of proton
decay suggests that discrete symmetries may play an important role
in the structure of the theory and, as Witten has recently
stressed \cite{Witten:2001bf}, such discrete symmetries might provide
an alternative explanation of the presence of massless doublets.

So even from the perspective of string theory, it is interesting
to explore conventional grand unification, with particular
emphasis on the role of discrete symmetries.  From this
perspective, however, there is a second puzzle: the origin of
$M_{\rm GUT}$. In field theory models of unification, this scale
often appears as an explicit input parameter.
One might expect that in a string theory
construction, the Higgs fields required to break the grand unified
group would be massless in some approximation.  In order that they
obtain large expectation values, they would need flat or nearly
flat potentials.  This need not necessarily be the
case -- after all, a parameter of only 10$^{-2}$
is required in order to obtain this scale from the Planck scale.
Still, the existence of a flat or nearly flat direction gives,
as we will see, an appealing picture, allowing for a natural mechanism
of generating the GUT scale in the effective field theory.
This possibility will be the focus of much of this paper.

We consider models in which $M_{\rm GUT}$ is given
by some combination of the Planck scale and the supersymmetry
breaking scale \cite{Babu:1994kb}. The GUT group, $G_{\rm GUT}$, in
our models, is broken by the 
expectation values of some fields, which correspond to an exact or
approximate flat direction in the limit of unbroken supersymmetry.
The location of the minimum, and the value of $M_{\rm GUT}$, are
ultimately determined by supersymmetry breaking effects.

It is easy to generate such an approximately flat direction
by the use of discrete symmetries. With symmetries such as a
$Z_{10}$, the potential along this direction arises from
non-renormalizable terms suppressed by a high power of the Planck
scale, such that, when balanced against a supersymmetry-breaking
soft mass for the GUT breaking field, the correct
value of $M_{\rm GUT}$ is obtained.

Alternatively, $M_{\rm GUT}$ may be an exact modulus of the theory.
This can easily be achieved with a continuous global symmetry
\cite{Hall:1995eq}, but such global symmetries are unlikely to arise
in a theory of gravity.  It is possible
to obtain  exact flat directions
with discrete symmetries, but typically one finds additional
light fields which spoil the predictions of unification.  We
explain these difficulties, and construct some examples of exact
flat directions which are almost, but not entirely, satisfactory.

The discrete symmetries of our models also play a role
in addressing the first puzzle mentioned above, namely,
the presence of colored partners of the Higgs doublets.
In a conventional grand unified model,
because of the GUT symmetry, the Yukawa couplings of
these partners are related by the gauge symmetry to
standard-model Yukawa couplings of quarks and leptons, giving rise
to fast proton decay. In the models we study, following ideas of
~\cite{Barbieri:1994jq,Barr:1996kp,Witten:2001bf},
a discrete symmetry distinguishes the triplets from the doublets.
This symmetry not only allows a triplet mass while forbidding
a doublet mass, but, as stressed in
\cite{Witten:2001bf}, may also suppress the triplet couplings
to matter fields relative to the usual Yukawa couplings.
Thus, proton decay is avoided by splitting not just the {\it masses}
but also the {\it couplings} of the doublets and triplets.

It is important to note that, as explained recently
in~\cite{Murayama:2001ur},
gauge coupling unification in minimal $SU(5)$ implies an allowed {\it window}
for the triplet masses, which is somewhat below $M_{\rm GUT}$.
Proton decay, on the other hand, implies a lower bound, which lies above
this window. The idea of Yukawa splitting between triplets and
doublets may avoid this problem in more general models. (For another
mechanism, see ref. \cite{Babu:2002fs}.)

As articulated by Witten, a symmetry that distinguishes doublets
from triplets requires a semi-simple group, and indeed our models
have $SU(5)\times SU(5)$ as $G_{\rm GUT}$. Still, the standard model
gauge group  lies in the diagonal $SU(5)$,
so that the main achievements of grand unification are
maintained: hypercharge is quantized, and all standard model couplings
originate from a single coupling, that of the diagonal $SU(5)$.
No large couplings are therefore required in order to ensure
unification. In addition, the models require relatively
few fields to achieve the initial stage
of symmetry breaking, so that they
remain weakly coupled up to the Planck scale.

We review the mechanism of doublet-triplet splitting by discrete
symmetries  of~\cite{Barbieri:1994jq,Barr:1996kp,Witten:2001bf}
in Section~\ref{doublettriplet} and describe the basic
$SU(5)\times SU(5)$ theory and flat direction we consider.
We explain how $M_{\rm GUT}$ may arise naturally along an approximate
flat direction in Section~\ref{gutscale}.
Some explicit models that realize these two ideas are
exhibited in Section~\ref{mesonsection}.
We argue that these are the simplest models which satisfy
all of our requirements: along this direction,
all these fields get $M_{\rm GUT}$ masses, and a discrete
symmetry, which forbids a doublet mass, remains unbroken. The models
contain two
pairs of bifundamentals, three adjoints and a gauge singlet.
As we show in Appendix~A, within such models a continuous
symmetry is required to obtain an exact flat direction.

In Section~\ref{baryonsection}, we show that an exact flat
direction, which is nearly satisfactory, can
be obtained by considering a different pattern of expectation values.
However, in these models, not
all fields with standard model quantum numbers gain mass.
In fact, a full $SU(5)$ adjoint representation remains
light, so that $SU(3)$ becomes strongly coupled near $M_{\rm GUT}$.

In the theories we consider, the standard-model Higgs
doublets may originate from a 5 and a $\bar5$ either of the same
$SU(5)$ factor, or of different $SU(5)$ factors.
To cancel $SU(5)$ anomalies, one can then arrange the
standard model matter fields in different representations
of $SU(5)\times SU(5)$, or add an extra $5+\bar 5$ pair.
We consider these different possibilities in Section~\ref{higgssection}.
It should be emphasized that the discrete symmetry
must be broken at some stage in order to allow a $\mu$ term
for the standard model Higgs doublets, and we discuss some possibilities
for accomplishing this.
In particular, we show that a Higgs doublet mass may be generated
automatically once the discrete symmetry is broken, without
having to add any new Higgs couplings.

Since, in this framework, the standard model matter and Higgs fields
may originate from representations of the different $SU(5)$'s,
some of their Yukawa couplings arise from non-renormalizable terms.
This situation introduces a small parameter of order $M_{\rm GUT}/M_{\rm Pl}
\sim10^{-2}$ into the Yukawa matrices. As we show in
Section~\ref{yukawasection} (with further details given in
Appendix~\ref{yukawaappendix}), the resulting quark and lepton mass
matrices are
viable. Furthermore, some of the mass hierarchies are automatically
explained by the gauge quantum numbers. When the standard Yukawa
couplings of the Higgs doublets arise from non-renormalizable terms,
the Yukawa couplings of of their colored partners can be highly
suppressed. This {\it Yukawa-splitting} is also explained in
Section~\ref{yukawasection}.

The framework described above, of an effective theory in which
$M_{\rm GUT}$ corresponds to an exact or approximate modulus, as we
have argued, seems a plausible outcome of string theory. String theory
typically leaves many light, undetermined moduli, and possesses
discrete symmetries. We discuss some distinctions between field theory
unification and string unification in Section~\ref{string}.

%%%%%%%%%
%%%%%%%%%
\section{Splitting triplets from doublets}
\label{doublettriplet}
In this section and the next, we describe the key elements of our models.
As explained in the introduction, we wish to distinguish
between Higgs doublets and triplets by a discrete symmetry.
(Here and in the following, we use ``Higgs'' to denote
the standard model doublets and their triplet partners,
as opposed to GUT breaking fields.) As discussed most clearly by Witten
\cite{Witten:2001bf}, if one wants to explain the
masslessness of Higgs doublets through discrete symmetries, there should
be an unbroken discrete symmetry (at least at a high energy
scale) which acts differently on doublets and triplets, and is
not a subgroup of hypercharge. This is not possible with a single $SU(5)$,
$SO(10)$ or $E_6$, but
it is not difficult to achieve with the group $SU(5) \times SU(5)$,
and other direct product groups, provided one has fields
transforming as bifundamentals. In this paper, we study the simplest such case,
with an $SU(5)\times SU(5)$ gauge group. Here, one wants a symmetry which
is a linear combination of a discrete symmetry acting on the bifundamentals,
and a gauge transformation acting in one of the $SU(5)$'s.  The latter
transformation can be taken to lie in the first $SU(5)$, without loss of
generality (by combining with a hypercharge transformation),
\beq\label{dishyp}
g_1 = \left ( \matrix{\alpha^{-1} &  &  &  &  \cr  &  \alpha^{-1}  & & &  \cr
  &  & \alpha^{-1} & &  \cr   &  & & \alpha^{{N+3 \over 2}} &  \cr  &    & & &
  \alpha^{{N+3 \over 2}}}\right )\ .
 \eeq
Here $\alpha$ is an $N$'th root of unity, and for simplicity we
have taken $N$ odd (for $N$ even it is a simple matter to
modify our formulas).

Suppose that one has two pairs of bifundamentals, $\Phi_i,\bar \Phi_i$,
$i=1,2$, and that the superpotential respects the $Z_N$ symmetry
\beq\label{origzn}
\Phi_1 \rightarrow \alpha \Phi_1,\ \ \
\bar \Phi_1 \rightarrow \alpha^{-1} \bar \Phi_1,\ \ \
\Phi_2 \rightarrow \alpha^{-{N+3 \over 2}} \Phi_2,\ \ \
\bar \Phi_2 \rightarrow \alpha^{{N+3 \over 2}} \bar \Phi_2\ .
\eeq
Then the vacuum expectation values (VEVs)
\beq\label{mesonvev}
\langle\Phi_1\rangle = \langle\bar\Phi_1\rangle =
 \left ( \matrix{v_1 &  &  &  &  \cr  &  v_1  & & &  \cr
  &  & v_1 & &  \cr   &  & & 0 &  \cr  &    & & & 0} \right ),\ \ \
\langle\Phi_2\rangle  = \langle\bar \Phi_2\rangle =
\left(\matrix{0 &  &  &  &  \cr  &  0  & & &  \cr
  &  & 0 & &  \cr   &  & & v_2 &  \cr  &    & & & v_2}\right ),
 \eeq
preserve a $Z_N^\prime$ symmetry which is a combination of the original $Z_N$
symmetry of eq. (\ref{origzn}) and the hypercharge transformation of eq.
(\ref{dishyp}).
Because the symmetry is unbroken, this structure of expectation values
is natural; it is automatically an extremum of the potential.
We will discuss another possible set of expectation values,
with only $\langle\Phi_1\rangle$ and $\langle\Phi_2\rangle$ nonzero, in
section~\ref{baryonsection}.

In fact, Barr constructed a model with precisely these features,
for the case $Z_N=Z_2$ \cite{Barr:1996kp}.
The superpotential in his model is:
\beqa
W &=& M_1 (\Phi_1 \bar \Phi_1+\Phi_2 \bar \Phi_2)\no\\
&+& {1\over M_2} \,\left[(\bar \Phi_1 \bar \Phi_1 \Phi_1 \bar \Phi_1)
+  (\Phi_2 \bar \Phi_2 \Phi_2 \bar \Phi_2)
+  (\Phi_1 \bar \Phi_1 \Phi_2 \bar \Phi_2)
+  (\Phi_1 \bar \Phi_2 \Phi_2 \bar \Phi_1)
\right. \no\\
&+& (\bar \Phi_1 \Phi_1)(\bar \Phi_1 \Phi_1)
+ (\bar \Phi_2 \Phi_2)(\bar \Phi_2 \Phi_2)
+ (\bar \Phi_1 \Phi_1)(\bar \Phi_2 \Phi_2)
+ (\bar \Phi_1 \Phi_2)(\bar \Phi_2 \Phi_1)\no\\
&+& \left. (\Phi_1 \bar \Phi_2 \Phi_1 \bar \Phi_2)
+  (\Phi_2 \bar \Phi_1 \Phi_2 \bar \Phi_1)
+ (\bar \Phi_1 \Phi_2)(\bar \Phi_1 \Phi_2)
+ (\bar \Phi_2 \Phi_1)(\bar \Phi_2 \Phi_1)\right].
\eeqa
(For simplicity, we have not distinguished the different couplings.)
It is easy to see that for such a superpotential, there is a solution
of the form~(\ref{mesonvev}).  Moreover, one can also see that
all fields gain mass.
Note, in particular, that the only continuous symmetry
of the superpotential is $SU(5) \times SU(5)$.  As a result, if we
turn on the gauge coupling, the only massless fields we expect to
find are those associated with the Goldstone bosons for the
breaking of this symmetry and their superpartners, which are Higgsed.

This superpotential is not of the form we
seek. The VEVs~(\ref{mesonvev}) are not a flat direction
of the superpotential. Instead, the potential has a minimum with VEVs
of order
$M_{\rm GUT}$ if $M_1$ and $M_2$ are of order $M_{\rm GUT}$.
$M_{\rm GUT}$ appears as an explicit input parameter;
both as mass terms ($M_1$), and as the scale suppressing
higher-dimension terms ($M_2$). This superpotential is puzzling if the
$\Phi$'s are the only $M_{\rm GUT}$ fields.
Generating these couplings in a theory with only fields of ${\cal
  O}(M_{\rm GUT})$-masses and marginal or relevant couplings requires
the addition of, at least, a $(24,24)$~\cite{Barr:1996kp}
with mass of order $M_{\rm GUT}$ and couplings to the $\Phi$ fields.

However, for the purposes of splitting doublets and triplets,
Barr's model is the simplest example of the mechanism we will use.
It is now clear how triplets can gain mass while doublets
remain massless. Assigning $Z_N$ charges to the Higgses so that they
cannot gain mass means that they can have couplings to $\Phi_1$ or
$\bar\Phi_1$,  but not to $\Phi_2$ or $\bar\Phi_2$. From the point of
view of the low energy theory,
it is easy to choose $Z_N$ charges for the Higgses, such
that the unbroken $Z_N^\prime$ will forbid a doublet mass.

If we want the unbroken $Z_N^\prime$ to distinguish between
doublet and triplet {\it mass terms}, the standard model Higgses
must arise from fields charged under different $SU(5)$
factors.\footnote{For $h(5,1)$ and $\bar h(\bar5,1)$,
the doublet and triplet mass terms $\bar h_3 h_3$
and $\bar h_2 h_2$ have the same $Z_N^\prime$ charges.}
The standard model Higgses therefore come from, for example,
\beq\label{smhiggs}
h(5, 1)\ ,\ \ \ \bar h^\prime(1,\bar5)\ .
\eeq
(Here and in the following, unprimed (primed) fields transform under
the first (second) $SU(5)$.) This possibility was discussed in
\cite{Witten:2001bf}. We refine this statement and consider additional
possibilities in Section~\ref{higgssection}.

The fact that the unbroken $Z_N^\prime$ symmetry in the low energy
theory below $M_{\rm GUT}$ distinguishes Higgs doublets and triplets
may be used to relax the bound on the triplet mass from
proton decay. Suppose that triplets are lighter than $M_{\rm GUT}$.
Since triplets and doublets have different charges under the unbroken
$Z_N^\prime$ of the low-energy theory, their couplings to
standard-model matter fields are typically not the same. From the
point of view of the full theory, if, as mentioned
above, the Higgs doublets come from a 5 and $\bar5$
of different $SU(5)$'s, then some standard-model Yukawa couplings
originate from higher-dimension terms, involving the bifundamentals.
Thus, the triplets couplings to matter fields may
be suppressed, so that their contribution to proton decay is small.
Of course, the triplet mass has to be sufficiently
high in order to preserve the unification of couplings.
But in minimal $SU(5)$ models, this requirement
actually gives both an {\sl upper} and a lower bound on the triplet mass.
While in our models the triplet masses arise from renormalizable
couplings to $\Phi_1$ and $\bar\Phi_1$, it may be possible
to construct models in which these masses are lower,
as low as permitted by gauge coupling unification.

%%%%%%%%%%%%%%%%%%%%%%%%%%%%%%%%%%%%%%%%%%%%%%%%%%%%%%%%
%%%%%%%%%%%%%%%%%%%%%%%%%%%%%%%%%%%%%%%%%%%%%%%%%%%%%%%%
\section{The origin of $M_{\rm GUT}$}
\label{gutscale}
%%%%%%%%%%%%%%%%%%%%%%%%%%%%%%%%%%%%%%%%%%%%%%%%%%%%%%%%
We are interested in models in which $M_{\rm GUT}$
corresponds to an exact, or nearly exact flat direction.
The main difficulty in constructing such models
is that, while no potential should be generated
along this direction (or at least no sizable potential),
all fields charged under the standard model gauge group
should acquire ${\cal O}(M_{\rm GUT})$ masses. As we will see, however,
it is possible to construct $SU(5)\times SU(5)$ models that
are fairly simple yet possess these two features.

We note that various constructions of this type have been put
forward in the past.  For example, a model based on $SO(10)$, which
also solves the doublet-triplet problem (using the so-called
``Dimopoulos-Wilczek mechanism") was presented
in \cite{Hall:1995eq}\footnote{Other examples have been given in
\cite{Agashe:1998kg,Cheng:1997fk,Graesser:1998ij}.
However, in most cases, these models were only studied at the level
of renormalizable terms.}. In this model, the existence of flat
directions, as well as the masslessness of the
Higgs doublets, are insured by global continuous $R$-symmetries, which
restrict the Lagrangian to renormalizable terms. The pattern
of expectation values which give massless doublets
does not respect any particular symmetry; it is an accident
of the restriction of the model to renormalizable terms. Allowing
only discrete symmetries, the flat direction is no longer flat,
and the desired pattern of expectation values requires fine tuning.
It should be noted that, even with the use of
global symmetries, the construction requires six adjoints, two
symmetric tensors and two spinor representations just in order to obtain
the first stage of symmetry breaking.

If one does not worry about the spectrum, it is not difficult to
obtain models in which the flat direction is an exact modulus
using discrete $R$ symmetries.  For example, suppose one has a
conventional $SU(5)$ theory with a single adjoint.  Requiring that
the adjoint, $A$, be neutral under the $R$ symmetry, while the
superpotential transforms with a non-trivial phase, forbids any
superpotential for $A$.  So the potential for $A$ is flat.
However, this simple example also illustrates the main difficulty:
if we choose to break to $SU(3)\times SU(2)\times U(1)$, there are
additional massless fields transforming as $(8,1)$ and $(1,3)$
under $SU(3) \times SU(2)$. This difficulty arises also in the
case of approximate flat directions. E.g., if one has an ordinary
(non-$R$) $Z_{10}$ under which $A\to e^{2\pi i\over 10} A$, then
the leading term in the superpotential involving $A$ alone is
$A^{10}$, so the flat direction is very flat, but the octet and
triplet are very light.

We will consider, as above, product groups. Most of our models
will involve two pairs of bifundamentals, as described in
section~\ref{mesonsection}, with the VEVs~(\ref{mesonvev}). To
avoid the problem of light fields, we will study models with
additional fields beyond the minimal set of bifundamentals. The
$Z_N$ global symmetry we discussed does not forbid a
superpotential in the would-be flat directions. In fact, all
nonzero gauge invariants, such as $\Phi_1\bar\Phi_1$,
$\Phi_2\bar\Phi_2$, and $\Phi_1^3\Phi_2^2$ (contracted with
$\varepsilon$-tensors) are also $Z_N$ singlets, and can appear in
the superpotential to any power. It is easy to forbid such terms
using an additional continuous symmetry in order to obtain an
exact flat direction. Discrete $R$-symmetries can give exact flat
directions involving the $\Phi$ fields but, as in our simple
$SU(5)$ example, there are unwanted massless fields in these
directions.  As we will show, with the simplest possible field
content, one cannot obtain exact flat directions with all fields
massive using only discrete symmetries (continuous global symmetries
can do the job).

But the fact that the flat directions are only approximate can be
a virtue, yielding a simple mechanism for generating $M_{\rm GUT}$
\cite{Babu:1994kb}. Suppose that the first contribution to the $F$
terms comes from the superpotential term
\beq\label{gutw}
W= {1\over M_{\rm Pl}^{n-3}} X^n\ ,
\eeq
where $X$ stands collectively for the
fields of the theory and $M_{\rm Pl}$ is the Planck scale. Suppose
further that, once supersymmetry is broken, $X$ acquires a
negative soft mass squared, $m^2$. Then the potential for $X$ is
\beq\label{gutv}
V= -m^2 \vert X\vert^2 + {1\over M_{\rm
    Pl}^{2n-6}} \vert X \vert ^{2n-2}\ ,
\eeq
leading to an $X$-VEV of the order
\beq\label{gutvalue}
\langle X\rangle\sim {\left({m\over
      M_{\rm Pl}}\right)}^{{1\over n-2}}\, M_{\rm Pl}\ .
\eeq
For $n$ around 10, this is around $M_{\rm GUT}$.

We therefore aim, in our models, for superpotentials
which generate $F$ terms only from nonrenormalizable
terms such as $(\Phi_1\bar\Phi_1)^5$, or $(\Phi_1^3\Phi_2^2)^2$.

In models in which $M_{\rm GUT}$ corresponds to an
exactly flat direction, we will assume that $M_{\rm GUT}$ is generated
by supersymmetry breaking effects.
Explicit examples of this type were studied
in~\cite{lutygmsb} in the context of gauge mediated
supersymmetry breaking, and in~\cite{lutyandus}
in the context of anomaly mediated supersymmtery breaking.
The mechanism of~\cite{lutyandus} leads however to
large flavor violation, unless extra structure is invoked.
We note that other mechanisms for generating $M_{\rm GUT}$ exist
in the literature~\cite{Cheng:1997fk,Graesser:1998ij}. In these
models, $M_{\rm GUT}$ is related to the dynamical scale of some
gauge group.

In the context of string theory, there is another mechanism which
one might imagine could generate $M_{\rm GUT}$.  This is the
appearance of a Fayet-Iliopoulos term, with a coefficient of order
$\alpha_{\rm GUT} \over 4 \pi$, perhaps small enough to explain
the scale \cite{Izawa:1999cm}. In this case, $M_{\rm GUT}$ might not
be related to supersymmetry breaking.  It is still necessary to have
directions in field space which are flat or nearly flat with respect
to the unified gauge group and the superpotential.

%%%%%%%%%%%%%%%%%%%%%%%%%%%%%%%%%%%%%%%%%%%%%%%%%
%%%%%%%%%%%%%%%%%%%%%%%%%%%%%%%%%%%%%%%%%%%%%
\section{Models with `Mesonic' Flat Directions}
\label{mesonsection}
We now turn to some examples which demonstrate
the various ideas we have discussed.
As described in the introduction, our models contain two pairs
of bifundamentals $\Phi_i$ and $\bar\Phi_i$,~$i=1,2$.
We impose a $Z_N$ global symmetry, which
ensures the masslessness of the Higgs doublets.
The $Z_N$ charges of the bifundamentals are listed
in table~\ref{tab:basicmeson}.
Note that the bifundamentals form vectorlike representations
of the $Z_N$ -- the charges of $\Phi_1$ and $\bar\Phi_1$ sum
to zero, and similarly the charges of $\Phi_2$ and $\bar\Phi_2$.

We would like to find a vanishing potential along the direction
discussed in the introduction [eq.~(\ref{mesonvev})].
Along this direction, $SU(5)\times SU(5)\times Z_N$ breaks into
$SU(3)_D\times SU(2)_D\times U(1)_D\times Z_N^\prime$.
$Z_N^\prime$ is a combination of the original discrete symmetry and a
discrete ``hypercharge'' subgroup of $SU(5)_1$.
Various components of $\Phi_i$, $\bar\Phi_i$ are uneaten.
In particular, the following standard-model $SU(3)\times SU(2)$
representations appear:
$3\times(8,1)$, $3\times(1,3)$, $2\times(3,2)$ and $2\times(\bar3,2)$.

It is easy to see that we cannot give mass to the
bifundamentals by couplings among the bifundamentals alone,
without spoiling the flat directions. Adding two fields transforming
as adjoints under the diagonal $SU(5)$ forces $Z_N$ charges that are
inconsistent with an exact low-energy $Z_N^\prime$ and, furthermore,
leaves light, incomplete $SU(5)$ multiplets. At a minimum, we need to
include in the model three fields transforming
as adjoints under the diagonal $SU(5)$.  We
add three adjoints of $SU(5)_1$, $A_{i=1,2,3}$,
as well as a gauge singlet $S$, with the superpotential
\beq\label{typicalw}
W=\lambda_{12}\Phi_1 A_1\bar\Phi_2+\lambda_{21}\Phi_2 A_2\bar\Phi_1+
\lambda_{11}\Phi_1 A_3\bar\Phi_1+\lambda_{22}\Phi_2 A_3\bar\Phi_2
+\eta_{12}SA_1A_2+\eta_{33}SA_3A_3\ .
\eeq
This superpotential indeed preserves a $Z_N$ symmetry,
for appropriate choices of the singlet and adjoint
charges.
The field content of the model is summarized in Table~\ref{tab:basicmeson}.
\begin{table}[ht]
\begin{center}
\begin{tabular}{cc} \hline \hline

Field           &  $SU(5)\times SU(5)\times Z_N$        \\ \hline\hline
$\Phi_1$        &  $(5,\bar5,1)$        \\
$\bar\Phi_1$    &  $(\bar5,5,-1)$       \\
$\Phi_2$        &  $(5,\bar5,(N-3)/2)$  \\
$\bar\Phi_2$    &  $(\bar5,5,(N+3)/2)$  \\
$A_1$           &  $(24,1,(N-5)/2)$     \\
$A_2$           &  $(24,1,(N+5)/2)$     \\
$A_3$           &  $(24,1,0)$           \\
$S$             &  $(1,1,0)$            \\ \hline
$h$             &  $(5,1,1)$            \\
$\bar h^\prime$ &  $(1,\bar5,0)$        \\ \hline
$\bar h$        &  $(\bar5,1,0)$        \\
$h^\prime$      &  $(1,5,-1)$           \\ \hline\hline
\end{tabular}
\end{center}
\caption{Basic fields and their charge assignments}
\label{tab:basicmeson}
\end{table}
The fields appearing below the first horizontal line contain the
Higgs doublets. Most often, these will come from $h$ and $\bar h^\prime$,
which we always include. The remaining two fields, $h^\prime$ and $\bar h$,
may or may not appear. We postpone further discussion of these Higgses until
Section~\ref{higgssection}, and for now concentrate on the GUT
breaking fields. In Appendix B we show that the field content in
Table~\ref{tab:basicmeson} is still small enough that there are no
Landau poles below the Planck scale.

Indeed, the potential vanishes for
\beqa\label{barvev}
\langle\Phi_1\rangle=\langle\bar\Phi_1\rangle&=&
v_1\times{\rm diag}(1,1,1,0,0),\no\\
\langle\Phi_2\rangle=\langle\bar\Phi_2\rangle&=&
v_2\times{\rm diag}(0,0,0,1,1),\no\\
\langle S\rangle&=&s,\\
\langle A_i\rangle\no&=&0, \eeqa and all pseudo-Goldstone bosons
get mass. The flatness condition requires\footnote{To see that
one needs to include Lagrnage multiplier terms ensuring the
tracelessness of the adjoints.} \beq\label{flatconb}
\lambda_{11}v_1^2=\lambda_{22}v_2^2. \eeq For superpotential
couplings of order one, the two VEVs are not very different.

Clearly, however, various terms which are allowed by the symmetries
would spoil the flatness of the potential. For example,
any power of $\Phi_1\bar\Phi_1$,  $\Phi_1^3\Phi_2^2$,
or $S$, generates a potential. In the following, we consider different
symmetries that forbid all or some of these terms.

%%%%%%%%%%%%%%%%%%%%%%%%%%%%%%%%%%%%%%%%%%%%%%%%%%%%
\subsection{Exact flat directions}
\label{exactsection}
%%%%%%%%%%%%%%%%%%%%%%%%%%%%%%%%%%%%%%%%%%%%%%%%%%%%
If one allows continuous global
symmetries, it is easy to obtain models in which the direction~(\ref{barvev})
is exactly flat. The superpotential~(\ref{typicalw}) preserves
a $U(1)_R$ symmetry under which all adjoints $A_i$ have charge $1$, the singlet
$S$ has charge $-1$, and the bifundamentals $\Phi_i$, $\bar\Phi_i$ have charge
zero. The superpotential has charge $1$.
For completeness we list the charges of all fields in Table~\ref{tab:exact}.
\begin{table}[ht]
\begin{center}
\begin{tabular}{cc} \hline \hline
Field           &  $SU(5)\times SU(5)\times Z_N\times U(1)_R$ \\ \hline\hline
$\Phi_1$        &  $(5,\bar5,1,0)$      \\
$\bar\Phi_1$    &  $(\bar5,5,N-1,0)$    \\
$\Phi_2$        &  $(5,\bar5,(N-3)/2,0)$        \\
$\bar\Phi_2$    &  $(\bar5,5,(N+3)/2,0)$        \\
$A_1$           &  $(24,1,(N-5)/2,1)$ \\
$A_2$           &  $(24,1,(N+5)/2,1)$ \\
$A_3$           &  $(24,1,0,1)$ \\
$S$             &  $(1,1,0,-1)$ \\  \hline\hline
\end{tabular}
\end{center}
\caption{Exact flat direction}
\label{tab:exact}
\end{table}

The $U(1)_R$ symmetry forbids all the `dangerous' non-renormalizable terms
that would otherwise lift the flat direction. It does so by requiring
that in each allowed term there is at least
one adjoint. All terms with two adjoints or with a single $A_1$ or $A_2$
are trivially harmless. The non-renormalizable terms with a single $A_3$
modify the flatness condition (\ref{flatconb}); the precise relation between
$v_1$ and $v_2$ is then changed, but not the fact that there is an exactly
flat direction of the form (\ref{mesonvev}).

If, however, we consider just a discrete $R$ symmetry, the
flat direction is lifted. The problem is easy to understand.
To obtain an exact flat direction for the $\Phi$'s,
we would like them to be neutral under some $R$ symmetry.
But this fixes the $R$ transformation laws of the $A_i$'s
and, as a result, of $S$. In particular, $S$ cannot
be neutral. We show in Appendix~A that there are always
terms of the form $S^n(\bar\Phi\Phi)^m$ which transform
properly under the symmetries.  So in this class of models, it is
easy to obtain an exact flat direction for the bifundamentals
using only discrete $R$ symmetries, but it is not possible to obtain
an exact flat direction simultaneously for $S$.  In the next
section, we discuss approximate flat directions.

%%%%%%%%%%%%%%%%%%%%%%%%%%%%%%%%%%%%%%%%%%%%%%%%%%%%%%%%
\subsection{Discrete Symmetry: Approximate flat directions}
\label{approxsection}
%%%%%%%%%%%%%%%%%%%%%%%%%%%%%%%%%%%%%%%%%%%%%%%%%%%%%%%%%%
As we just saw, we can get a model with an approximate
flat direction by replacing the continuous $R$ symmetry above
by a discrete $R$ symmetry, $Z_M^{R}$. (The field content and charge
assignments of this model are presented explicitly in Table
\ref{tab:approx}.) Then the terms
\beq
{1\over M_{\rm Pl}^{M-4}}S^{M-1} \ ,\ \ \
{1\over M_{\rm Pl}^{M-2}}S^{M-1} \Phi_i\bar\Phi_i,\ \ i=1,2,
\eeq
are allowed and lift the flat direction.

In this case, the $S$ flat direction is lifted by
a superpotential term of dimension $M-1$, while the $\Phi$
flat direction (here $\Phi$ stands for all bifundamentals)
is lifted by superpotential terms of dimension $M+1$.
As discussed in Section~\ref{gutscale}, if the singlet and bifundamentals
get negative  masses-squared by supersymmetry-breaking,
they will be stabilized near $M_{\rm GUT}$ for $M\sim 9,10$.

We thus obtain a model in which the GUT breaking VEVs
are naturally obtained at the right scale, all fields
associated with the GUT breaking obtain masses of order $M_{\rm GUT}$,
and the low energy theory has an unbroken global $Z_N^\prime$
symmetry, which can distinguish between Higgs doublets and triplets.

It is straightforward to generate other models that share these features.
For example, consider giving the $\Phi$'s
nonzero  charges under the $R$ symmetry. We take the $R$ symmetry to
be  $Z_{11}^{R}$ with the charges given in Table~\ref{tab:modelB}.
\begin{table}[ht]
\begin{center}
\begin{tabular}{cc} \hline \hline
Field           &  $SU(5)\times SU(5)\times Z_N\times Z_{11}^R$ \\ \hline\hline
$\Phi_1$        &  $(5,\bar5,1,0)$      \\
$\bar\Phi_1$    &  $(\bar5,5,N-1,0)$    \\
$\Phi_2$        &  $(5,\bar5,(N-3)/2,3)$        \\
$\bar\Phi_2$    &  $(\bar5,5,(N+3)/2,8)$        \\
$A_1$           &  $(24,1,(N-5)/2,4)$ \\
$A_2$           &  $(24,1,(N+5)/2,9)$ \\
$A_3$           &  $(24,1,0,1)$ \\
$S$             &  $(1,1,0,10)$ \\ \hline\hline
\end{tabular}
\end{center}
\caption{Model B: Approximate flat direction}
\label{tab:modelB}
\end{table}
The symmetries allow, among others, the following terms:
\beq\label{allbarj}
{1\over M_{\rm Pl}^6}S^4(\bar\Phi_1^3\bar\Phi_2^2)\ ,\ \ \
{1\over M_{\rm Pl}^7}(\Phi_1^3\Phi_2^2)^2.
\eeq
The flat direction is lifted by terms of dimensions 9 and 10
and consequently a reasonable value for $M_{\rm GUT}$ is obtained.

%%%%%%%%%%%%%%%%%%%%%%%%%%%%%%%%%%%%%%%%%%%%%%%%%
%%%%%%%%%%%%%%%%%%%%%%%%%%%%%%%%%%%%%%%%%%%%%%%
\section{Models with `baryonic' flat directions}
\label{baryonsection}
%%%%%%%%%%%%%%%%%%%%%%%%%%%%%%%%%%%%%%%%%%%%%%%%%
As we saw in the previous section, it is impossible
in the models considered so far to obtain an exact mesonic flat direction
using discrete symmetries only. This can easily be done, however,
if we consider the following direction:
\beqa\label{baryvev}
\langle\Phi_1\rangle&=&v\times{\rm diag}(1,1,1,0,0),\no\\
\langle\Phi_2\rangle&=&v\times{\rm diag}(0,0,0,1,1),\no\\
\langle\bar\Phi_i\rangle&=&0.
\eeqa
Again, these VEVs break the gauge symmetry to the
standard-model gauge group, and preserve the $Z_N^\prime$
global symmetry irrespectively of the $\bar\Phi_i$
charge assignments [the $Z_N$-charges of $\Phi_1$ and $\Phi_2$
are still $1$ and $(N-3)/2$].

Apart from the possibility of obtaining exact flat directions,
these directions are also promising for obtaining potentials
which are flat to a very high degree.  The only
non-vanishing gauge invariant combinations are powers of the
baryon operator $\Phi_1^3\Phi_2^2$. If the first such term
appearing in the superpotential is the baryon squared, it can
naturally generate the scale $M_{\rm GUT}$ once supersymmetry is
broken.

It is easy to see that discrete symmetries can guarantee
an exactly vanishing potential in this case.
For example, with no additional fields in the model,
we can choose a $Z_M^{R}$ symmetry under which the $\Phi$'s
have charge zero and the $\bar\Phi$'s have charge
one. Then, assuming the superpotential carries $R$-charge 2,
the only terms allowed contain two $\bar\Phi$'s,
and the potential vanishes.

It is, however, very hard to generate masses for all
GUT breaking fields in these models,
or even for all fields but one (or more) sets in a complete $SU(5)$ representation.
In particular, note that the uneaten fields in the $\Phi$'s
are an $SU(3)$ octet and an $SU(2)$ triplet,
and these can only get mass from the `baryon' operators
$\Phi_1^5$ and $\Phi_1\Phi_2^4$.

In the following, we present three types of models. Each of these
models demonstrates some nice features that can be achieved with
`baryonic' flat directions, but also suffers from some problems. Model
A is closest in spirit to the models with `mesonic' flat directions
discussed in the previous section. It has an approximately flat potential
but a light 24 of $SU(5)$. With such a large representation, the $SU(3)$
coupling blows up below $M_{\rm GUT}$, unless the fields in the 24
get masses of at least 1~TeV. Model B has no light fields beyond
the MSSM fields, but the flat direction is lifted by a dimension-5
superpotential term. Model C has an exact flat direction but, again,
has a light 24. Both models B and C have no adjoints, but in addition
to the light 24 have incomplete multiplets at a scale of order
$10^{12}$ GeV which might spoil coupling unification.

%%%%%%%%
\subsection{Model A: Approximate Flat Direction and a Light 24}
The field content of our first model appears in
Table~\ref{tab:baryon}. We take the superpotential to carry $R$-charge
1.
\begin{table}[ht]
\begin{center}
\begin{tabular}{cc} \hline \hline
Field           &  $SU(5)\times SU(5)\times Z_N\times Z_9^R$\\ \hline\hline
$\Phi_1$        &  $(5,\bar5,1,1)$      \\
$\bar\Phi_1$    &  $(\bar5,5,N-1,0)$    \\
$\Phi_2$        &  $(5,\bar5,(N-3)/2,1)$        \\
$\bar\Phi_2$    &  $(\bar5,5,(N+3)/2,0)$        \\
$A$             &  $(24,1,0,0)$ \\
$B_1$           &  $(1,24,(N+5)/2,0)$ \\
$B_2$           &  $(1,24,(N-5)/2,0)$ \\
$S_1$             &  $(1,1,-5,0)$ \\
$S_2$             &  $(1,1,5,0)$ \\
\hline\hline
\end{tabular}
\end{center}
\caption{`Baryon' model A}
\label{tab:baryon}
\end{table}
The symmetry allows the superpotential
\beqa\label{wbar1}
W&=& B_2 \Phi_1\bar\Phi_2 + B_1\Phi_2 \bar\Phi_1 + A \Phi_1\bar\Phi_1
+A\Phi_2\bar\Phi_2 + S_1 B_1^2 + S_2 B_2^2\no\\
&+& {1\over M_{\rm Pl}^7}(\Phi_1^3 \Phi_2^2)^2
+ {1\over M_{\rm Pl}^9} (S_1 S_2) (\Phi_1^3 \Phi_2^2)^2 +\cdots .
\eeqa
Here we are interested in nonzero VEVs for $S_1$, $S_2$ as well.
We only show terms that either give mass to some fields,
or generate a potential. We get an approximate flat direction,
allowing for a natural generation of $M_{\rm GUT}$.
Note, however, that for general $N$, the $Z_N$ symmetry
is spontaneously broken by the singlet VEVs, and one needs
to carefully choose the Higgs charges, so that the doublets remain light.
In addition, one needs to ensure that higher-dimension terms
involving the bifundamentals and the singlets,
that destabilize the VEV~(\ref{baryvev}), are sufficiently suppressed.
Alternatively, if we insist on having an unbroken
global symmetry, we can work with $N=2$, with only $\Phi_2$,
$\bar\Phi_2$, and $B_i$ odd under the $Z_2$.
The unpleasant feature of this model is that it leaves a light 24.

%%%%%%%%
\subsection{Model B: Approximate Flat Direction with No Light Fields}
We can write down a much simpler model with just the $\Phi$'s and $\bar\Phi$'s.
We impose a $Z_2 \times Z_2$ global symmetry. The first $Z_2$ is our
original $Z_N$ with $N=2$, under which only $\Phi_2$ and $\bar\Phi_2$ are odd.
Under the second $Z_2$ the two $\bar\Phi$'s are odd.
The allowed superpotential is:
\beqa\label{wbar2}
W&=&{1\over M}\left(\Phi_1 \bar\Phi_1 \Phi_1 \bar\Phi_1 +
\Phi_1 \bar\Phi_2 \Phi_1 \bar\Phi_2 +
\Phi_2 \bar\Phi_1 \Phi_2 \bar\Phi_1 +
\Phi_2 \bar\Phi_2 \Phi_2 \bar\Phi_2 +
\Phi_1 \bar\Phi_1 \Phi_2 \bar\Phi_2 +
\Phi_1 \bar\Phi_2 \Phi_2 \bar\Phi_1\right)\no\\
&+&{1\over M^2}\left( \Phi_1^5 + \Phi_1\Phi_2^4 + \Phi_1^3
  \Phi_2^2\right) + \cdots\ .
\eeqa
The dimension-4 terms give mass to all components of
the $\bar\Phi$'s (these therefore get mass two orders
of magnitude below $M_{\rm GUT}$).
The  first two dimension-5 terms give mass to the $SU(2)$-triplet
and $SU(3)$-octet in the $\Phi$'s (these therefore get mass
four orders of magnitude below $M_{\rm GUT}$, a
potential problem for coupling unification).
Finally, the problematic term is the last one; it lifts
the flat direction already at dimension-5, and is thus
inappropriate for generating $M_{\rm GUT}$. In order that the model is
acceptable, we need to identify $M=M_{\rm GUT}$ in eq. (\ref{wbar2}),
that is, put $M_{\rm GUT}$ by hand.

%%%%%%%%
\subsection{Model C: Exactly Flat Direction}
%%%%%%%%
Consider a model with the bifundamentals and singlets, where the
$\bar\Phi_i$ fields carry a $Z_M^R$-charge $+1$ and all other fields
($\Phi_i$ and $S_i$) are $Z_M^R$-neutral. The superpotential has
$Z_M^R$-charge $+2$. Thus all terms must include two
$\bar\Phi_i$-fields and consequently the baryonic direction
(\ref{baryvev}) is exactly flat. Consider additional symmetries that
allow only the following terms up to dimension-5:
\beq\label{supoiii}
W={1\over M_{\rm Pl}^2}\left(S_1\Phi_1\bar\Phi_1\Phi_1\bar\Phi_2
  +S_2\Phi_2\bar\Phi_1\Phi_2\bar\Phi_2
+S_3\Phi_1\bar\Phi_1\Phi_2\bar\Phi_1\right).
\eeq
This superpotential leaves a light 24, while all other fields [which
include $(8,1)+(1,3)$ plus complete $SU(5)$-multiplets] get their
masses at a scale of order $M_{\rm GUT}^3/M_{\rm Pl}^2$ which may be
problematic for coupling unification.

%%%%%%%%%%%%%%%%%%%%%%%%%%%%%%%%%%%%%%%%%%%%%%
%%%%%%%%%%%%%%%%%%%%%%%%%%%%%%%%%%%%%%%%%%%
\section{Standard model Higgses}
\label{higgssection}
%%%%%%%%%%%%%%%%%%%%%%%%%%%%%%%%%%%%%%%%%
%%%%%%%%%%%%%%%%%%%%%%%%%%%%%%%%%%%%%%%%%%%
So far, we have concentrated on the GUT breaking fields.
In Section~\ref{mesonsection}, we found models in
which $M_{\rm GUT}$ corresponds to an exact or approximate
flat direction. Along this direction, all fields associated with the
GUT breaking are heavy. In addition, there is an unbroken $Z^\prime_N$
global symmetry. We will now see in more detail how this symmetry
splits the doublets and triplets.
For concreteness, we focus on the first model described
in Section~\ref{approxsection}, but it is straightforward
to repeat this discussion for different versions of the
model -- only the charges under the discrete $R$-symmetry will change.
For convenience, we list in Table \ref{tab:approx} the relevant
fields.
\begin{table}[ht]
\begin{center}
\begin{tabular}{cc} \hline \hline
Field           &  $SU(5)\times SU(5)\times Z_N\times Z_M^R$ \\ \hline\hline
$\Phi_1$        &  $(5,\bar5,1,0)$      \\
$\bar\Phi_1$    &  $(\bar5,5,N-1,0)$    \\
$\Phi_2$        &  $(5,\bar5,(N-3)/2,0)$        \\
$\bar\Phi_2$    &  $(\bar5,5,(N+3)/2,0)$        \\
$A_1$           &  $(24,1,(N-5)/2,1)$ \\
$A_2$           &  $(24,1,(N+5)/2,1)$ \\
$A_3$           &  $(24,1,0,1)$ \\
$S$             &  $(1,1,0,M-1)$ \\ \hline
$h$             &  $(5,1,1,1)$          \\
$\bar h^\prime$ &  $(1,\bar5,0,0)$      \\ \hline
$\bar h$        &  $(\bar5,1,0,0)$      \\

$h^\prime$      &  $(1,5,N-1,1)$                \\
\hline\hline
\end{tabular}
\end{center}
\caption{The fields and symmetries of the model of Section \ref{approxsection}}
\label{tab:approx}
\end{table}

Let us next write down the most general renormalizable
superpotential that involves
the $h$ fields:
\beq\label{suupoh}
W_1=h\bar\Phi_1\bar h^\prime+h^\prime\Phi_1\bar h.
\eeq
Since the $h$ fields do not couple to the $\Phi_2$ and $S$ fields, only the
triplets acquire masses.

It is useful to look at the charges of the doublets and triplets
under the unbroken $Z_N^\prime$ of the low energy theory.
We list these in Table~\ref{tab:higgs}.
\begin{table}[ht]
\begin{center}
\begin{tabular}{cc} \hline \hline
Field           &  $Z_N^\prime$ \\ \hline
$h_3$           &  $0$          \\
$h_2$           &  $(N+5)/2$            \\
$\bar h_3$      &  $1$          \\
$\bar h_2$      &  $(N-3)/2$            \\
$h^\prime_3$    &  $-1$         \\
$h^\prime_2$    &  $-1$         \\
$\bar h^\prime_3$& $0$          \\
$\bar h^\prime_2$& $0$          \\
\hline\hline
\end{tabular}
\end{center}
\caption{Doublet ($h_2$) and triplet ($h_3$) $Z_N^\prime$-charges.}
\label{tab:higgs}
\end{table}
Clearly, no doublet mass term is allowed in the low energy theory
below $M_{\rm GUT}$.
Therefore, if we start from four Higgses, the theory contains four
light doublets.

It is easy to see that this result always holds if we insist on an unbroken
$Z_N^\prime$ and masses of order $M_{\rm GUT}$ for all triplets. In
order to give mass to
all triplets, we need to allow both terms in $W_1$ of eq. (\ref{suupoh}).
Since $\Phi_1$ and $\bar\Phi_1$ have opposite $Z_N$ charges,
the $Z_N$ charges of the Higgses must satisfy
\beq
Q(h) +Q(\bar h^\prime) +  Q(h^\prime) + Q(\bar h) =0 \ .
\eeq
As a result, the terms
\beq
h\bar\Phi_2\bar h^\prime\ ,\ \ h^\prime \Phi_2 \bar h\ ,
\eeq
have the same $Z_N$ charges: they are either both allowed,
or both forbidden. We are therefore left with four light doublets.
There are then three options a priori:
\begin{itemize}
\item {\bf A}: The theory does not contain $h^\prime$ and $\bar h$.
$SU(5)$ anomalies are cancelled by appropriate choices of $SU(5)\times SU(5)$
representations for the standard model matter fields.
The standard model Higgses come from fields charged under different $SU(5)$'s.
Some standard model Yukawa couplings arise from non-renormalizable terms.
$Z_N^\prime$ can be broken by supersymmetry breaking effects
to generate the $\mu$ term.
\item {\bf B}: The theory does contain $h^\prime$ and $\bar h$,
but these remain massless\footnote{It is easy to forbid
the relevant mass terms by choosing appropriate charges
for  $h^\prime$ and $\bar h$.}~\cite{Witten:2001bf}.
Witten speculates that these could be the messengers
of supersymmetry breaking.
The heavy triplets are from $h$ and $\bar h^\prime$.
\begin{itemize}
\item {\bf B1}: The standard model Higgses come from fields charged under
different $SU(5)$'s, say, $h$ and $\bar h^\prime$, so that, again,
some standard model Yukawa couplings arise from non-renormalizable terms.
\item {\bf B2}: The standard model Higgses come from fields charged under
a single $SU(5)$, say, $h$ and $\bar h$. Then, standard model fields can all
be charged under the same $SU(5)$, and all Yukawa couplings are renormalizable.
\end{itemize}
\item {\bf C}: The theory does contain $h^\prime$ and $\bar h$.
All triplets gain mass through the couplings~(\ref{suupoh}).
The $Z_N^\prime$ is broken at a high scale,  so that one
doublet pair also gets mass around $M_{\rm GUT}$.
It is possible to arrange for an acceptable $\mu$
term for the remaining two doublets, for example, through the
mechanisms proposed in \cite{Leurer:1993gy} or in \cite{Barr:1996kp}.
This is most easily done by adding a gauge-singlet, $S_H$,
charged under the $Z_N$, with a GUT scale VEV. Then, in order to have
masses of order $M_{\rm GUT}$,
the doublets that couple to $S_H$ must be charged under the
same $SU(5)$, and the higgs doublets are charged under the second $SU(5)$.

In this case, one should also make sure that the pattern of
VEVs we are considering, eqn.~(\ref{mesonvev}),
is not destabilized, that is, that
$\Phi_1$, $\bar\Phi_1$ do not get VEVs in the last two entries.
It is easy to see that such VEVs can be very small.
For example, if $S_H$ has charge 1 under the $Z_N$,
then the operator
$S_H^{(N+5)/2} S^{M-1} \bar\Phi_1 \Phi_2 \bar\Phi_1\Phi_1$
is allowed, and can generate the danegrous VEV.
However, it is very suppressed. In fact, such operators
might even generate a $\mu$ term of precisely the right
size, without having to add any new coupling for the Higgses.
The doublet mass will simply arise from the first term
in~(\ref{suupoh}).\footnote{With a $Z_2$ instead of a $Z_N$, one
does not even have to break the $Z_2$ in order to
generate a large mass for the extra triplets. This is
the mechanism employed in~\cite{Barr:1996kp}.}
\end{itemize}

In the next section, we will study the implications
for quark and lepton mass matrices.

%%%%%%%%%%%%%%%%%%%%%%%%%%%%%%%%%%%%%%%%%%%%%%%%%%%%%%%%%%
%%%%%%%%%%%%%%%%%%%%%%%%%%%%%%%%%%%%%%%%%%%%%%%%%%%%%
\section{Yukawa Matrices and Yukawa Splitting}
\label{yukawasection}
%%%%%%%%%%%%%%%%%%%%%%%%%%%%%%%%%%%%%%%%%%%%%%%%%%%%
We now turn to the Yukawa couplings for quarks and leptons.
We will see that some of the possibilities mentioned above
are excluded, and that in others, the $SU(5)\times SU(5)$
gauge symmetry automatically suppresses some Yukawa couplings.
Moreover, in many cases, triplet-Higgses couplings to matter
are naturally suppressed,
demonstrating the mechanism of ``Yukawa splitting''.

The chiral part of our model contains the Higgs and matter fields. Taking into
account the various possibilities for the Higgs fields discussed in the
previous section and
the requirement of anomaly cancellation, we have the following options:

{\bf A:} Standard model higgses from $(5,1)_H+(1,\bar5)_H$, no additional
fundamentals, and the standard model fermion generations coming from
one of the following sets of representations:
  \begin{enumerate}
  \item $3\times(\bar5,1)+2\times(10,1)+(1,10)$
  \item $2\times[(\bar5,1)+(1,10)]+(10,1)+(1,\bar5)$
  \item $(\bar5,1)+2\times(1,\bar5)+3\times(1,10)$
  \end{enumerate}

{\bf B1:} Standard model higgses from $(5,1)_H+(1,\bar5)_H$, additional
fundamentals in $(\bar5,1)_H+(1,5)_H$, and the standard model fermion
generations coming from one of the following sets of representations:
  \begin{enumerate}
  \item $3\times[(\bar5,1)+(10,1)]$

  \item $2\times[(\bar5,1)+(10,1)]+(1,\bar5)+(1,10)$
  \item $(\bar5,1)+(10,1)+2\times[(1,\bar5)+(1,10)]$
  \item $3\times[(1,\bar5)+(1,10)]$
  \end{enumerate}

{\bf B2, C:} Standard model higgses from $(5,1)_H+(\bar5,1)_H$,
additional fundamentals in $(1,\bar5)_H+(1,5)_H$, and the standard model
fermion generations coming from one of the following sets of representations:
  \begin{enumerate}
  \item $3\times[(\bar5,1)+(10,1)]$
  \item $2\times[(\bar5,1)+(10,1)]+(1,\bar5)+(1,10)$
  \item $(\bar5,1)+(10,1)+2\times[(1,\bar5)+(1,10)]$
  \item $3\times[(1,\bar5)+(1,10)]$
  \end{enumerate}

Note that the only renormalizable Yukawa couplings involve
\beq\label{renyuk}
(10,1)(10,1)(5,1)_H,\ \ \ (1,10)(1,\bar5)(1,\bar5)_H,\ \ \
(10,1)(\bar5,1)(\bar5,1)_H.
\eeq
Non-renormalizable terms will involve one or two bifundamental fields. The
resulting Yukawa couplings are suppressed therefore by one or two powers
of $\epsilon=M_{\rm GUT}/M_{\rm Pl}\sim10^{-2}$.
This leads to several interesting consequences.

First, a phenomenologically viable model must have a renormalizable top-Yukawa.
This requirement excludes model (3) of class {\bf A} and models (4) of classes
{\bf B} and {\bf C}.

Second, there is a single possibility to have all Yukawa couplings coming
from renormalizable terms, and that is model (1) of classes {\bf B2} and
{\bf C}. Here the gauge symmetry plays no role in the Yukawa hierarchy.

Third, when down-type Yukawas invlove matter fields
charged under different $SU(5)$'s, say, a $(10,1)$ and a
$(1,\bar5)$, they give {\it either} a down-quark 
{\it or} a charged-lepton mass term.
This allows for the exciting possibility that,
just by the gauge quantum numbers, bottom-tau unification
is maintained, but similar relations for the two light
generations are not!

Furthermore, we see that it is possible to explain some
features of the Yukawa hierarchy
by the $SU(5)\times SU(5)$ symmetry. Take, for example, model (1) of class
{\bf B1}. The mass matrices here are of the form
\beqa\label{parsupoa}
M_u\sim\langle(5,1)_H\rangle\pmatrix{1&1&1\cr1&1&1\cr1&1&1\cr},\ \ \
M_d\sim\langle(1,\bar5)_H\rangle\pmatrix{\epsilon&\epsilon&\epsilon\cr
 \epsilon&\epsilon&\epsilon\cr \epsilon&\epsilon&\epsilon\cr}.
\eeqa
Thus, the small ratio $m_b/m_t$ is explained by the gauge symmetry.
The hierarchy within each sector requires, however, some additional
flavor physics.
We show the mass matrices obtained in the remaining cases in
Appendix~\ref{yukawaappendix}.

Finally, we can see how doublet-triplet {\it Yukawa-splitting}
works. Take again model (1) of class {\bf B1}. The entries in $M_d$ of
eq. (\ref{parsupoa}) come from the non-renormalizable terms
\beq\label{noremd}
(10,1)(\bar5,1)(1,\bar5)_H(\bar5,5)_{\bar\Phi_2}.
\eeq
Note that these terms induce Yukawa couplings for the doublet Higgs fields
only. Yukawa couplings for the triplet Higgs fields could have been induced
by the following terms:
\beq\label{noretr}
(10,1)(\bar5,1)(1,\bar5)_H(\bar5,5)_{\bar\Phi_1}.
\eeq
However, if the discrete charge assignments are such that the terms in
(\ref{noremd}) are allowed, those in (\ref{noretr}) are, in general,
forbidden. This situation can be understood also in terms of the unbroken
low-energy $Z_N^\prime$ symmetry that can allow Yukawa couplings for the
doublets and forbid them for the triplets. Thus, the triplet-Yukawa couplings
will only appear with $Z_N^\prime$-breaking and can therefore be very
strongly suppressed. As explained in section \ref{doublettriplet}, this
mechanism of doublet-triplet Yukawa splitting is very helpful
in solving the problem of proton decay.

%%%%%%%%%%%%%%%%%%%%%%%%%%%%%%%%%%%%%%%%%%%%%%%%
%%%%%%%%%%%%%%%%%%%%%%%%%%%%%%%%%%%%%%%%%%%%%
\section{Conclusions:  Distinctions Between Grand Unification and String Unification}
\label{string}
%%%%%%%%%%%%%%%%%%%%%%%%%%%%%%%%%%%%%%%%%%%%%%%
%%%%%%%%%%%%%%%%%%%%%%%%%%%%%%%%%%%%%%%%%%%%%%%%
Various puzzles of grand unification find attractive solutions in
models that combine product gauge groups and discrete symmetries.
In this work we focused on a framework that combines an
$SU(5)\times SU(5)$ gauge group with Abelian ($Z_N$) discrete
symmetries.

The gauge group breaks directly, at a scale $M_{\rm GUT}$, into the
standard model $SU(3)\times SU(2)\times U(1)$ group which resides in
the diagonal $SU(5)$. Let us first emphasize again that, in spite of
the fact that there is no range of energies where there is a simple
unifying gauge group, the attractive features of GUTs are maintained:
there is coupling unification and hypercharge is quantized.
At the same time, some of the problems associated with GUTs find
solutions for which
the combination of product groups and discrete symmetries is essential:

(i) The doublet-triplet splitting problem has a symmetry-based
solution. A discrete symmetry (that remains unbroken below
$M_{\rm GUT}$) can forbid doublet masses while allowing triplet
masses. At the same time, this symmetry can allow Yukawa couplings
for the doublet while forbidding them for the triplet.
Thus the inconsistency between the constraints from coupling unification
and from proton decay that arises in minimal $SU(5)$ can be
avoided in this framework.

(ii) The GUT scale corresponds to an exact or an approximate flat
direction. In the latter case, the value of $M_{\rm GUT}$ is not
an input parameter but is induced by an interplay between the
Planck scale and the supersymmetry breaking scale.  In addition to
its intrinsic appeal, we argued that
if the underlying theory is string or M theory, it is almost
inevitable that $M_{\rm GUT}$ should be a modulus.

The combination of product gauge groups and discrete
symmetries might have interesting implications for a third puzzle:

(iii) A non-trivial flavor structure arises since, in general,
some of the Yukawa couplings arise from non-renormalizable terms.
Such couplings are suppressed therefore by powers of
$M_{\rm GUT}/M_{\rm Pl}$.

Our models demonstrate that it is possible to have fully
consistent models of grand unification, with no light, extra
incomplete (or even complete) $SU(5)$ multiplets and with a
relatively simple particle content at $M_{\rm GUT}$, and in which
the value of $M_{\rm GUT}$ is explained in terms of the Planck
scale and the supersymmetry breaking scale.

The assumption that $M_{\rm GUT}$ is a modulus opens up the
possibility of a cosmological moduli problem.  In the context of
string theory, the difficulty is that moduli typically dominate
the energy density at nucleosynthesis.  Unless they are very
massive, their subsequent decays reheat the universe only to
temperatures of order a few keV, and the successful predictions of
big bang nucleosynthesis are spoiled
\cite{Banks:1993en,deCarlos:1993jw}.  In the present case,
however, the couplings of the modulus may be barely strong enough
to evade this problem.  Explicitly, the coefficient $c$ in the
leading contribution to the decay amplitude, ${\cal L}_{\cal
M}={c\alpha_{\rm GUT}\over4\pi M_{\rm GUT}}{\cal M}FF$, can
be ${\cal O}(50)$ due to the large number of fields.

In the introduction, we stressed that while string theory does not
admit continuous global symmetries, it often yields discrete
symmetries.  These symmetries are usually gauge symmetries, and
are subject to various consistency conditions. In weakly coupled
string models as well as a range of strongly coupled string
theories \cite{Banks:1991xj,Dine:1993zf,dg,Witten:2001bf}, it is
possible to cancel all anomalies in discrete symmetries by
postulating a discrete transformation law for some moduli fields.
A priori, however, it is not clear that this provides a constraint
on field theory models, since one might postulate the existence of
several such moduli with Planck scale couplings.  It is true that
in most instances which have been studied, a single field can
cancel all anomalies, but it seems unlikely that this holds in
general \cite{dg}.  We have not explored possible anomaly
constraints in our work, but this is certainly an issue worthy of
further study.

Whether or not of stringy origin, unification in the framework of
product groups opens up new possibilities regarding the flavor structure
of quarks and leptons and the question of proton decay. We have
presented some aspects of the quark Yukawa hierarchy and of the
doublet-triplet Yukawa splitting in Section~\ref{yukawasection}.
We are currently pursuing further investigations of these intriguing
possibilities.

%%%%%%%%%%%%
\begin{acknowledgments}
We thank Michael Graesser, Graham Kribs, Lisa Randall, Guy Raz and
Yuri Shirman for useful discussions.
The visit of MD to the Weizmann Institute was supported by the Albert Einstein
Minerva Center for Theoretical Physics and by the United States - Israel
Binational Science Foundation (BSF). The work of MD is supported in
part by the US Department of Energy.
YN\ is supported by the Israel Science Foundation founded by the
Israel Academy of Sciences and Humanities.
\end{acknowledgments}

%%%%%%%%%%%%%%%%%%%%%%%%%%%%%%%%%%%%%%%%%%%%%%%%%%
%%%%%%%%%%%%%%%%%%%%%%%%%%%%%%%%%%%%%%%%%%%%%%%
\appendix
\section{Exact flat directions}
%%%%%%%%%%%%%%%%%%%%%%%%%%%%%%%%%%%%%%%%%%%%%%%
%%%%%%%%%%%%%%%%%%%%%%%%%%%%%%%%%%%%%%%%%%%%%%%%%
In this appendix we show that it is impossible to forbid all
higher-dimension operators using a $Z_N\times Z_M^{R}$ discrete symmetry
in a model with bifundamentals, adjoints and singlets.

We denote the $Z_M^{R}$-charge of each field by the corresponding small
letter. For example, the charge of $\Phi_i$ is $\phi_i$. The superpotential
has charge $w$. All the equations involving the $Z_M^R$-charges are
mod $M$, and all charges are integer.

In order to allow the terms $ S A_3^2$ and $\Phi_1 A_3 \bar\Phi_1$
of eq. (\ref{typicalw}), the charges should satisfy
\beq
\phi_1 + \bar\phi_1 +a_3 =w\ ,\ \ \ s+2a_3 = w\ .
\eeq
We will now show that we can always find $n$, $m$ so that
\beq
\label{allow}
S^n {\left(\Phi_1\bar\Phi_1\right)}^m \
\eeq
is allowed. For this term to appear we need:
\beq
n(w-2a_3) + m(w-a_3)=w\ \Longrightarrow\ (n+m-1)w=(2n+m)a_3.
\eeq
This condition is satisfied for
\beq\label{mequation}
n+m-1=q M \ ,\ \ 2n+m=q^\prime M\ ,
\eeq
where $q$, $q^\prime$ are some integers. Now consider the charges under
the $Z_N$ symmetry. We can repeat the above, with $w\to 0$ and $M\to N$.
Therefore we get one additional equation for $n$, $m$:
\beq\label{nequation}
2n+m=q^{\prime\prime} N\ .
\eeq
The three equations~(\ref{mequation})-(\ref{nequation})
can be solved with $q^\prime=q^{\prime\prime}=k N$,
with integer $k$, giving:
\beq
n=k N M - q M -1\ ,\ \ \ m=(2q-k N)M+2\ .
\eeq
Since the charges are only defined mod($M N$) we can choose:
\beq
n=MN - qM -1\ ,\ \ \ m=2q M+2\ .
\eeq
A simple possibility is $q=0$: the term
\beq\label{soln}
S^{NM-1} {\left(\Phi_1\bar\Phi_1\right)}^2
\eeq
is always allowed.

Note that if we have effectively just the $Z_M^{R}$ symmetry,
(as in the models of section \ref{mesonsection}, where both $S$ and
$\Phi_i\bar\Phi_i$ are neutral under the $Z_N$), then
\beq
S^{M-1} {\left(\Phi_1\bar\Phi_1\right)}^2
\eeq
is allowed (corresponding to $N=1$ in~(\ref{soln})).

%%%%%%%%%%%%%%%%%%%%%%%%%%%%%%%%%%%%%%%%%%%%
%%%%%%%%%%%%%%%%%%%%%%%%%%%%%%%%%%%%%%%%%%%%
\section{Landau Poles}
%%%%%%%%%%%%%%%%%%%%%%%%%%%%%%%%%%%%%%%%%%%%%
%%%%%%%%%%%%%%%%%%%%%%%%%%%%%%%%%%%%%%%%%%%%%
Since our models require a large number of representations at $M_{\rm GUT}$,
there is a potential problem of hitting a Landau pole below the Planck scale.
We investigate this issue here. The RGE reads
\beq\label{rgeeq}
{1\over\alpha(M_1)}={1\over\alpha(M_2)}+{b\over2\pi}\ln{M_2\over M_1}.
\eeq
We will hit a Landau pole, $\alpha\to\infty$, at a scale $M_{\rm LP}$ given by
\beq\label{lanpol}
{M_{\rm LP}\over M}=\exp\left({2\pi\over b\alpha(M)}\right).
\eeq
Using $\alpha(M_{\rm GUT})\sim0.04$ and requiring that the Landau pole is
not reached below the Planck scale, that is $M_{\rm LP}/M_{\rm GUT}\gsim 10^2$,
we find $b\lsim 34$.

For a supersymmetric $SU(N)$ group, we have
\beqa\label{defbfun}
b&=&\sum_a C(\phi_a)-3C({\rm adj}),\no\\
C({\rm fun})&=&1/2,\ \ \ C({\rm adj})=N,\ \ \ C({\rm as})=N/2-1,
\eeqa
where fun, adj and as stand for, respectively, the fundamental, adjoint and
antisymmetric representations. Defining for an $SU(5)\times SU(5)^\prime$
theory
\beqa\label{defnums}
n_h&=&\#(5,1)+\#(\bar5,1),\no\\
n_\Phi&=&\#(5,\bar5)+\#(\bar5,5),\no\\
n_A&=&\#(24,1),\no\\
n_T&=&\#(10,1)+\#(\overline{10},1),
\eeqa
we obtain for $b(SU(5))$:
\beq\label{bsufive}
b={1\over2}n_h+{5\over2}n_\Phi+5n_A+{3\over2}n_T-15.
\eeq

In particular, the maximal $b$ arises if we all the MSSM fields
($n_h=5$, $n_T=3$) and all $M_{gut}$ fields ($n_\Phi=4$, $n_A=3$)
transform under the same $SU(5)$:
\beq\label{favmod}
n_h=5,\ \ \ n_\Phi=4,\ \ \ n_A=3,\ \ \ n_T=3\ \ \ \Longrightarrow\ \ \ b=17.
\eeq
Putting that in eq. (\ref{lanpol}), with $\alpha_{\rm GUT}\simeq1/25$, we find
\beq\label{lapofive}
{M_{\rm LP}\over M_{\rm GUT}}\simeq\exp\left({50\pi\over17}\right)
\simeq10^4.
\eeq
Thus our models are safe against Landau poles.

%%%%%%%%%%%%%%%%%%%%%%%%%%%%
\section{Quark and Lepton masses}
\label{yukawaappendix}
In this Appendix, we give expressions for the
Yukawa matrices that arise in the various models we consider.
The representations for the quark (or lepton) fields are given
explicitly. The Higgs fields are denoted by $h(5,1)$, $\bar
h(\bar5,1)$ and $\bar h^\prime(1,\bar5)$. Bifundamental fields are
denoted by $\Phi(5,\bar5)$ and $\bar\Phi(\bar5,5)$.

For the up sector, the following combinations are relevant for inducing
the Yukawa couplings:
\beqa\label{upentries}
&(10,1)&(10,1)\ h,\no\\
&(1,10)&(1,10)\ h\ \bar\Phi,\no\\
&(10,1)&(1,10)\ h\ \Phi\ \Phi.
\eeqa
For the down and charged lepton sectors, the following combinations
are relevant for inducing the Yukawa couplings from $\bar h^\prime$:
\beqa\label{downentries}
&(1,10)&(1,\bar5)\ \bar h^\prime,\no\\
&(10,1)&(\bar5,1)\ \bar h^\prime\ \bar\Phi,\no\\
&(1,10)&(\bar5,1)\ \bar h^\prime\ \Phi\ \ \ [d/\ell],\no\\
&(10,1)&(1,\bar5)\ \bar h^\prime\ \bar\Phi\ \bar\Phi\ \ \ [d/\ell],
\eeqa
and from $\bar h$:
\beqa\label{dooentries}
&(10,1)&(\bar5,1)\ \bar h,\no\\
&(1,10)&(1,\bar5)\ \bar h\ \Phi,\no\\
&(10,1)&(1,\bar5)\ \bar h\ \bar\Phi\ \ \ [d/\ell],\no\\
&(1,10)&(\bar5,1)\ \bar h\ \Phi\ \Phi\ \ \ [d/\ell].
\eeqa
Entries marked with $[d,\ell]$ mean that these terms can give
masses to either the down sector or the charged lepton sector but not
to both. In the full high energy theory this depends on whether the
bifundamental representations are $\Phi_1$ or $\Phi_2$ (or
$\bar\Phi_1$ or $\bar\Phi_2$). From
the point of view of the low
energy theory, below $M_{\rm GUT}$, this depends on the
different $Z_N^\prime$ charges carried by $\bar d_L d_R$ and
$\bar\ell_R\ell_L$. ($Z_N^\prime$-breaking will eventually lift the
zero masses.)

Defining $\epsilon=M_{\rm GUT}/M_{\rm Pl}$, we get the following
paramatric suppression due to the $SU(5)\times SU(5)$ gauge symmetry
(entries in parenthesis vanish in either $M_d$ or $M_\ell$):

Model {\bf A}(1), with $(1,10)$ as first generation:
\beqa\label{parsupta}
M_u\sim\langle h\rangle\pmatrix{\epsilon&\epsilon^2&\epsilon^2\cr
 \epsilon ^2&1&1\cr \epsilon^2&1&1\cr},\ \ \
M_d\sim\langle\bar h^\prime\rangle\pmatrix{(\epsilon)&(\epsilon)&(\epsilon)\cr
 \epsilon&\epsilon&\epsilon\cr \epsilon&\epsilon&\epsilon\cr}.
\eeqa

Model {\bf A}(2), with $(10,1)$ as third generation and $(\bar5,1)$ as first
generation:
\beqa\label{parsuptb}
M_u\sim\langle h\rangle\pmatrix{\epsilon&\epsilon&\epsilon^2\cr
  \epsilon&\epsilon&\epsilon^2\cr \epsilon^2&\epsilon^2&1\cr},\ \ \
M_d\sim\langle\bar h^\prime\rangle\pmatrix{1&(\epsilon)&(\epsilon)\cr
  1&(\epsilon)&(\epsilon)\cr (\epsilon^2)&\epsilon&\epsilon\cr}.
\eeqa

Model {\bf B1}(1):
\beqa\label{parsupoaa}
M_u\sim\langle h\rangle\pmatrix{1&1&1\cr1&1&1\cr1&1&1\cr},\ \ \
M_d\sim\langle\bar h^\prime\rangle\pmatrix{\epsilon&\epsilon&\epsilon\cr
  \epsilon&\epsilon&\epsilon\cr \epsilon&\epsilon&\epsilon\cr}.
\eeqa

Model {\bf B1}(2), with $(1,10)+(1,\bar5)$ as first generation:
\beqa\label{parsupob}
M_u\sim\langle h\rangle\pmatrix{\epsilon&\epsilon^2&\epsilon^2\cr
  \epsilon^2&1&1\cr \epsilon^2&1&1\cr},\ \ \
M_d\sim\langle\bar h^\prime\rangle\pmatrix{1&(\epsilon)&(\epsilon)\cr
  (\epsilon^2)&\epsilon&\epsilon\cr (\epsilon^2)&\epsilon&\epsilon\cr}.
\eeqa

Model {\bf B1}(3), with $(10,1)+(\bar5,1)$ as third generation:
\beqa\label{parsupoc}
M_u\sim\langle h\rangle\pmatrix{\epsilon&\epsilon&\epsilon^2\cr
  \epsilon&\epsilon&\epsilon^2\cr \epsilon^2&\epsilon^2&1\cr},\ \ \
M_d\sim\langle\bar h^\prime\rangle\pmatrix{1&1&(\epsilon)\cr
  1&1&(\epsilon)\cr (\epsilon^2)&(\epsilon^2)&\epsilon\cr}.
\eeqa

Model {\bf B2/C}(1):
\beq\label{parsuoa}
M_u\sim\langle h\rangle\pmatrix{1&1&1\cr1&1&1\cr1&1&1\cr},\ \ \
M_d\sim\langle\bar h\rangle\pmatrix{1&1&1\cr 1&1&1\cr 1&1&1\cr}.
\eeq

Model {\bf B2/C}(2), with $(1,10)+(1,\bar5)$ as first generation:
\beq\label{parsuob}
M_u\sim\langle h\rangle\pmatrix{\epsilon&\epsilon^2&\epsilon^2\cr
  \epsilon^2&1&1\cr \epsilon^2&1&1\cr},\ \ \
M_d\sim\langle\bar h\rangle\pmatrix{\epsilon&(\epsilon^2)&(\epsilon^2)\cr
  (\epsilon)&1&1\cr (\epsilon)&1&1\cr}.
\eeq

Model {\bf B2/C}(3), with $(10,1)+(\bar5,1)$ as third generation:
\beq\label{parsuoc}
M_u\sim\langle h\rangle\pmatrix{\epsilon&\epsilon&\epsilon^2\cr
  \epsilon&\epsilon&\epsilon^2\cr \epsilon^2&\epsilon^2&1\cr},\ \ \
M_d\sim\langle\bar h\rangle\pmatrix{\epsilon&\epsilon&(\epsilon^2)\cr
  \epsilon&\epsilon&(\epsilon^2)\cr (\epsilon)&(\epsilon)&1\cr}.
\eeq

Note that, in principle, the $Z_N$ symmetry may constrain these
matrices further. We checked that none of the above models is excluded
by these constraints.

%%%%%%%%%%%%%%%%%%%%%

%%%%%%%%%%%%%%%%%%%%%%%%%%%%%%%
%%%%%%%%%%%%%%%%%%%%%%%%%

\begin{thebibliography}{01}

%\cite{Green:sp}
\bibitem{Green:sp}
M.~B.~Green, J.~H.~Schwarz and E.~Witten,
``Superstring Theory,''
{\it  Cambridge, Uk: Univ. Pr. (1987) (Cambridge Monographs
  On Mathematical Physics)}.

%\cite{Banks:1988yz}
\bibitem{Banks:1988yz}
T.~Banks and L.~J.~Dixon,
%``Constraints On String Vacua With Space-Time Supersymmetry,''
Nucl.\ Phys.\ B {\bf 307}, 93 (1988).
%%CITATION = NUPHA,B307,93;%%

%\cite{Witten:2001bf}
\bibitem{Witten:2001bf}
E.~Witten,
%``Deconstruction, G(2) holonomy, and doublet-triplet splitting,''
arXiv:hep-ph/0201018.
%%CITATION = HEP-PH 0201018;%%

%\cite{Babu:1994kb}
\bibitem{Babu:1994kb}
K.~S.~Babu and R.~N.~Mohapatra,
%``Mass matrix textures from superstring inspired SO(10) models,''
Phys.\ Rev.\ Lett.\  {\bf 74}, 2418 (1995)
[arXiv:hep-ph/9410326].
%%CITATION = HEP-PH 9410326;%%

%\cite{Hall:1995eq}
\bibitem{Hall:1995eq}
L.~J.~Hall and S.~Raby,
%``A Complete supersymmetric SO(10) model,''
Phys.\ Rev.\ D {\bf 51}, 6524 (1995)
[arXiv:hep-ph/9501298].
%%CITATION = HEP-PH 9501298;%%

%\cite{Barbieri:1994jq}
\bibitem{Barbieri:1994jq}
R.~Barbieri, G.~R.~Dvali and A.~Strumia,
%``Strings versus supersymmetric GUTs: Can they be reconciled?,''
Phys.\ Lett.\ B {\bf 333}, 79 (1994)
[arXiv:hep-ph/9404278].
%%CITATION = HEP-PH 9404278;%%

%\cite{Barr:1996kp}
\bibitem{Barr:1996kp}
S.~M.~Barr,
%``The stability of the gauge hierarchy in SU(5) x SU(5),''
Phys.\ Rev.\ D {\bf 55}, 6775 (1997)
[arXiv:hep-ph/9607359].
%%CITATION = HEP-PH 9607359;%%

%\cite{Murayama:2001ur}
\bibitem{Murayama:2001ur}
H.~Murayama and A.~Pierce,
%``Not even decoupling can save minimal supersymmetric SU(5),''
Phys.\ Rev.\ D {\bf 65}, 055009 (2002)
[arXiv:hep-ph/0108104].
%%CITATION = HEP-PH 0108104;%%

%\cite{Babu:2002fs}
\bibitem{Babu:2002fs}
K.~S.~Babu and S.~M.~Barr,
%``Eliminating the d = 5 proton decay operators from SUSY GUTs,''
Phys.\ Rev.\ D {\bf 65}, 095009 (2002)
[arXiv:hep-ph/0201130].
%%CITATION = HEP-PH 0201130;%%

%\cite{Agashe:1998kg}
\bibitem{Agashe:1998kg}
K.~Agashe,
%``GUT and SUSY breaking by the same field,''
Phys.\ Lett.\ B {\bf 444}, 61 (1998)
[arXiv:hep-ph/9809421].
%%CITATION = HEP-PH 9809421;%%

%\cite{Cheng:1997fk}
\bibitem{Cheng:1997fk}
H.~C.~Cheng,
%``Supersymmetric dynamical generation of the grand unification scale,''
Phys.\ Lett.\ B {\bf 410}, 45 (1997)
[arXiv:hep-ph/9702214].
%%CITATION = HEP-PH 9702214;%%

%\cite{Graesser:1998ij}
\bibitem{Graesser:1998ij}
M.~Graesser,
%``Getting the supersymmetric unification scale from quantum confinement  with chiral symmetry breaking,''
Phys.\ Rev.\ D {\bf 59}, 035007 (1999)
[arXiv:hep-ph/9805417].
%%CITATION = HEP-PH 9805417;%%

%\cite{Chacko:1998uu}
\bibitem{lutygmsb}
Z.~Chacko, M.~A.~Luty and E.~Ponton,
%``Dynamical determination of the unification scale by gauge-mediated
%supersymmetry breaking,''
Phys.\ Rev.\ D {\bf 59}, 035004 (1999)
[arXiv:hep-ph/9806398].
%%CITATION = HEP-PH 9806398;%%

%\cite{Chacko:2000wq}
\bibitem{lutyandus}
Z.~Chacko, M.~A.~Luty, E.~Ponton, Y.~Shadmi and Y.~Shirman,
%``The GUT scale and superpartner masses from anomaly mediated  supersymmetry breaking,''
Phys.\ Rev.\ D {\bf 64}, 055009 (2001)
[arXiv:hep-ph/0006047].
%%CITATION = HEP-PH 0006047;%%

%\cite{Izawa:1999cm}
\bibitem{Izawa:1999cm}
K.~I.~Izawa, K.~Kurosawa, Y.~Nomura and T.~Yanagida,
%``Grand-unification scale generation through the anomalous U(1) breaking,''
Phys.\ Rev.\ D {\bf 60}, 115016 (1999)
[arXiv:hep-ph/9904303].
%%CITATION = HEP-PH 9904303;%%

%\cite{Leurer:1993gy}
\bibitem{Leurer:1993gy}
M.~Leurer, Y.~Nir and N.~Seiberg,
%``Mass matrix models: The Sequel,''
Nucl.\ Phys.\ B {\bf 420}, 468 (1994)
[arXiv:hep-ph/9310320].
%%CITATION = HEP-PH 9310320;%%

%\cite{Banks:1993en}
\bibitem{Banks:1993en}
T.~Banks, D.~B.~Kaplan and A.~E.~Nelson,
%``Cosmological implications of dynamical supersymmetry breaking,''
Phys.\ Rev.\ D {\bf 49}, 779 (1994)
[arXiv:hep-ph/9308292].
%%CITATION = HEP-PH 9308292;%%

%\cite{deCarlos:1993jw}
\bibitem{deCarlos:1993jw}
B.~de Carlos, J.~A.~Casas, F.~Quevedo and E.~Roulet,
%``Model independent properties and cosmological implications of the dilaton and moduli sectors of 4-d strings,''
Phys.\ Lett.\ B {\bf 318}, 447 (1993)
[arXiv:hep-ph/9308325].
%%CITATION = HEP-PH 9308325;%%

%\cite{Banks:1991xj}
\bibitem{Banks:1991xj}
T.~Banks and M.~Dine,
%``Note on discrete gauge anomalies,''
Phys.\ Rev.\ D {\bf 45}, 1424 (1992)
[arXiv:hep-th/9109045].
%%CITATION = HEP-TH 9109045;%%

%\cite{Dine:1993zf}
\bibitem{Dine:1993zf}
M.~Dine, R.~G.~Leigh and D.~A.~MacIntire,
%``Discrete gauge anomalies in string theory,''
arXiv:hep-th/9307152.
%%CITATION = HEP-TH 9307152;%%

\bibitem{dg}
M. Dine and M. Graesser, in preparation.

\end{thebibliography}
\end{document}